\documentclass[preprin,11pt]{elsarticle}
\usepackage[top=3cm,bottom=3.4cm,left=2cm,right=2cm]{geometry}

\makeatletter
\def\ps@pprintTitle{%
  \let\@oddhead\@empty
  \let\@evenhead\@empty
  \let\@oddfoot\@empty
  \let\@evenfoot\@oddfoot
}
\usepackage{xcolor}
\usepackage{bm}
\usepackage{hyperref}
\usepackage{amsmath}
\usepackage{amssymb}
\DeclareFontFamily{OT1}{pzc}{}
\DeclareFontShape{OT1}{pzc}{m}{it}{<-> s * [1.10] pzcmi7t}{}
\DeclareMathAlphabet{\mathpzc}{OT1}{pzc}{m}{it}
\newcommand{\nn}{\nonumber\\}

\newcommand{\ket}[1]{\left| #1 \right>\!}

\numberwithin{equation}{section}
\begin{document}
\begin{frontmatter}

\title{
Dirac-Born-Infeld action from inequivalent representations of open strings
}

\author[socu]{Isao Kishimoto}
\ead{ikishimo@rs.socu.ac.jp}

\author[nwu]{Tomohiko Takahashi}
\ead{tomo@asuka.phys.nara-wu.ac.jp}

\author[nwu]{Ayane Tezuka}
\ead{tezuka@asuka.phys.nara-wu.ac.jp}

\affiliation[socu]{
  organization={Center for Liberal Arts and Sciences,
  Sanyo-Onoda City University},
  addressline={Daigakudori 1-1-1},
  city={Sanyo-Onoda},
  state={Yamaguchi},
  postcode={756-0884},
  country={Japan}
}

\affiliation[nwu]{
  organization={Department of Physics, Nara Women's University},
  city={Nara},
  postcode={630-8506},
  country={Japan}
}

\begin{abstract}
We study open strings in a constant magnetic field from the viewpoint of
universal canonical variables. Using a conserved Klein-Gordon-type
pairing, we show that the canonical commutation relations remain
background independent in the distributional sense, while the background
dependence is encoded in the mode representation.  The Bogoliubov
transformation between different magnetic backgrounds is not unitarily
implementable, so that the corresponding Fock representations are
inequivalent. Nevertheless, the transformation determines the relative
normalization of the vacua.
We show that
the resulting relative vacuum normalization precisely reproduces
the relative boundary $g$-function, 
or equivalently the
magnetic-field-dependent part 
of the disk amplitude, and yields
the Dirac-Born-Infeld factor.  Thus the disk normalization is encoded
directly in the transformation between inequivalent representations of
universal open-string variables.
\end{abstract}

%\begin{keyword}
%Open string; Background independence; Bogoliubov transformation;
%Inequivalent representations; Dirac-Born-Infeld action
%\end{keyword}

\end{frontmatter}

%%%%%%%%%%%%%%%%%%%%%%%%%%%%%%%%%%%%%%%%%%%%%%%%%%%%%%
%\renewcommand{\thepage}{\arabic{page}}
%\renewcommand{\thefootnote}{\arabic{footnote}}
%\setcounter{page}{1}
%\setcounter{footnote}{0}
%\baselineskip=19pt plus 0.2pt minus 0.1pt
%%%%%%%%%%%%%%%%%%%%%%%%%%%%%%%%%%%%%%%%%%%%%%%%%%%%%%

%\tableofcontents

%%%%%%%%%%%%%%%%%%%%%%%%%%%%%%%%%%%%%%%%%%%%%%%%%%%%%%
\section{Introduction}

A nonperturbative formulation of string theory is expected
to describe different string backgrounds within a single theoretical
framework. An important idea toward this goal was proposed by
Kugo and Zwiebach~\cite{Kugo:1992md}:
the canonical string coordinates
$X^\mu(\sigma)$ and momenta $P_\mu(\sigma)$ may be regarded
as universal objects, independent of the background, while the
background dependence is encoded in their mode expansions and
in the corresponding representations of the canonical commutation
relations. Different backgrounds may therefore correspond to
inequivalent representations of the same universal canonical
variables, rather than to different states in a single Fock space.

The state space underlying string field theory has in fact been
a subtle issue since early studies of background-changing solutions.
Classical solutions describing dilaton condensation in light-cone
string field theory~\cite{Yoneya:1987gc} and background changes in the
$\alpha=p^+$ HIKKO theory~\cite{Kugo:1992md}
have singular coefficients when
represented in the original Fock basis, a behavior reminiscent of
a finite transformation between inequivalent representations.
Related state-space subtleties also appear in analytic solutions in
open string
field theory. The phantom term in Schnabl's solution~\cite{Schnabl:2005gv}
and the homotopy analysis~\cite{Inatomi:2011xr}
of identity-based solutions~\cite{Takahashi:2002ez,Kishimoto:2002xi} provide
examples in which states outside the conventional Fock-space
representation play an essential role in physical quantities or
BRST cohomology. These observations suggest that the appropriate
state space for nonperturbative vacua and background changes may
extend beyond a single Fock representation.

For open strings, a constant magnetic field provides a particularly
nontrivial test of this idea. The boundary conditions depend on the
magnetic field, and the string coordinates become noncommutative
at the endpoints~\cite{Abouelsaood:1986gd,Seiberg:1999vs}.
At first sight, this seems to imply that
the canonical commutation relations themselves depend on the
background, apparently conflicting with the universal-object
picture. Indeed, an earlier study of open string field theory in a
constant $B$-field background~\cite{Kawano:1999fw}
concluded from the endpoint
noncommutativity that the original string coordinates could not be
regarded as universal objects, and introduced modified coordinates
whose canonical commutation relations are background independent.
This apparent conflict with the universal-object picture motivates
the present work.

There is another aspect of the constant magnetic background that
makes this problem particularly instructive. In our previous study
of the corresponding open string field
theory solution~\cite{Ishibashi:2016xak},
the boundary $g$-function was determined from correlation functions
of boundary-condition-changing operators. The resulting disk
partition function contains the factor
$\sqrt{\det(1+2\pi\alpha' F)}$,
and hence reproduces the Dirac-Born-Infeld (DBI) action. Thus,
from the BCFT viewpoint, the DBI factor measures the change in the
physical normalization associated with the boundary condition.
This raises a natural question from the universal-object viewpoint:
can the same normalization be recovered directly from the change
of representation of the canonical open-string variables?

In this letter, we answer this question affirmatively.
We revisit the canonical quantization of an open string in a constant magnetic
field while keeping the canonical coordinates and momenta as
background-independent universal objects.
We show that the endpoint
noncommutativity does not modify the canonical commutation relations
when the latter are understood in the distributional sense, so that
the background dependence can be assigned entirely to the mode
expansion. We then construct the Bogoliubov transformation between
mode representations corresponding to different magnetic fields.
Its off-diagonal part fails to satisfy the Hilbert-Schmidt condition,
and therefore the corresponding Fock representations are unitarily
inequivalent.

Nevertheless, the Bogoliubov transformation retains information
that is not captured by identifying state vectors in a common Fock
space. Its infinitesimal form determines the relative normalization
of the vacua belonging to different representations.
We show that the resulting relative vacuum normalization
reproduces the relative boundary $g$-function,
or equivalently the magnetic field dependence of the disk normalization.
Integrating its variation gives the DBI factor. The same
quantity that was obtained in~\cite{Ishibashi:2016xak}
from the factorization of
boundary-condition-changing correlation functions thus emerges here
from the transformation between inequivalent representations of the
universal canonical algebra.

This provides an operator interpretation of the DBI normalization
that is complementary to its world-sheet path-integral
derivation~\cite{Fradkin:1985qd}.
Rather than evaluating the disk partition function directly,
we recover its background dependence from the Bogoliubov transformation
of the vacuum. In this sense, the disk normalization is already
encoded in how different string backgrounds are represented on the
universal open-string canonical variables.

\section{Open strings in background magnetic fields}

The canonical quantization of open strings in a constant background
gauge field, including the associated
mode expansion, was developed in~\cite{Abouelsaood:1986gd}.
Here we specialize this
construction to the neutral-string sector relevant to a constant
background field on a D-brane, using the notation and
parametrization of~\cite{Ishibashi:2016xak}.

Our purpose here is not to rederive the fixed-background quantization,
but to organize it in terms of the conserved pairing so that mode
representations at different backgrounds can be compared directly.

We consider bosonic string theory in 26-dimensional
Minkowski space-time with a D$p$-brane, imposing Neumann
boundary conditions on $X^\mu$ ($\mu=0,\cdots,p$) and Dirichlet
boundary conditions on $X^i$ ($i=p+1,\cdots,25$).
For a constant magnetic field $F_{\mu\nu}$ along the Neumann
directions, the gauge potential can be chosen as
\begin{align}
 A_\mu(X)=-\frac{1}{2}F_{\mu\nu}X^\nu.
\end{align}
Since $F_{\mu\nu}$ is a real antisymmetric tensor with $F_{0\mu}=0$,
it can be brought  into block-diagonal form by a
rotation in the spatial Neumann directions.
We therefore focus on a two-dimensional block
spanned by  $X^1$ and $X^2$ and introduce
complex coordinates, $X=(X^1+iX^2)/\sqrt{2}$ and 
$\tilde{X}=(X^1-iX^2)/\sqrt{2}$.
These variables satisfy the wave equation together with
the boundary conditions
\begin{align}
&
 X'+i\tan \pi\lambda\,\dot{X}=0\quad(\sigma=0,\pi),
\label{eq:bc1}
\\
&
 \tilde{X}'-i\tan \pi\lambda\,\dot{\tilde{X}}=0\quad(\sigma=0,\pi),
\label{eq:bc2}
\end{align}
where $\tan\pi\lambda=2\pi\alpha' F_{12}$.

Following~\cite{Abouelsaood:1986gd}, we use a conserved
Klein-Gordon-type sesquilinear form on the space of solutions.
The differential operator $D$, including its boundary contribution,
was introduced
in~\cite{Abouelsaood:1986gd}.
Here the conserved pairing will serve not only to
normalize the modes at a fixed background,
but also to extract the background-dependent mode coefficients
from the canonical string variables.
This will allow us to compare the mode representations associated
with different values of the background field. We define
\begin{align}
 (f,\,g)\equiv&
 \int_0^\pi f^*(t,\sigma) D g(t,\sigma) d\sigma
\end{align}
where  $D=i\stackrel{\rightarrow}{\partial}_t-
i\stackrel{\leftarrow}{\partial}_t
+\tan\pi\lambda\{\delta(\sigma)-\delta(\pi-\sigma)\}$.
Here, $f(t,\sigma)$ and $g(t,\sigma)$ satisfy  the wave equation with
\eqref{eq:bc1}.
Their complex conjugates
$f^*(t,\sigma)$ and $g^*(t,\sigma)$ satisfy
\eqref{eq:bc2}.
We use the same notation for the corresponding sesquilinear form between
solutions
satisfying \eqref{eq:bc2}.
\begin{align}
 (f^*,\,g^*)\equiv&
 \int_0^\pi f(t,\sigma) \tilde{D} g^*(t,\sigma) d\sigma
\nn
=&
-  \int_0^\pi g^*(t,\sigma) D f(t,\sigma) d\sigma,
\end{align}
where $\tilde{D}=-D^*=i\stackrel{\rightarrow}{\partial}_t-
i\stackrel{\leftarrow}{\partial}_t
-\tan\pi\lambda\{\delta(\sigma)-\delta(\pi-\sigma)\}$.
With this convention,
complex conjugation gives
\begin{align}
 (f,\,g)^*=(g,\,f)=-(f^*,g^*).
\end{align}
An important property of this form is
that it is independent of $t$,
even though the integration is performed only over
the spatial coordinate $\sigma$:
\begin{align}
 \frac{d}{dt}(f,g)=\frac{d}{dt}(f^*,g^*)=0.
\end{align}
Thus the mode coefficients extracted by this pairing are independent of
the time slice on which the pairing is evaluated.

We introduce the following set of mode
functions adapted to the two boundary
conditions. For \eqref{eq:bc1}, they are
\begin{align}
 \varphi_0(t,\sigma)=&
t \cos \pi\lambda-i\left(\sigma-\frac{\pi}{2}\right) \sin\pi \lambda,
\\
\psi_n(t,\sigma)=& e^{-int}\cos(n\sigma+\pi\lambda)\quad (n\in \mathbb{Z}).
\end{align}
For \eqref{eq:bc2}, we define the corresponding mode functions by
\begin{align}
\tilde{\varphi}_0(t,\sigma)
\equiv\varphi_0^* (t,\sigma),
\quad
\tilde{\psi}_n(t,\sigma)
\equiv\psi_{-n}^* (t,\sigma).
\end{align}
With the above definitions of the pairings,
the two sets of basis functions have the same normalization:
\begin{align}
&(\psi_n,\varphi_0)
=(\tilde{\psi}_n,\,\tilde{\varphi}_0)
=i\pi\delta_{n,0},
\label{eq:psiphi_1}\\
&
(\varphi_0,\psi_n)
=(\tilde{\varphi}_0,\,\tilde{\psi}_n)
=-i\pi\delta_{0,n},
\label{eq:psiphi_2}\\
&
(\psi_n,\psi_m)
=(\tilde{\psi}_n,\,\tilde{\psi}_m)
=n\pi\delta_{n,m}.
\label{eq:psipsi}
\end{align}
The form is indefinite, reflecting the underlying symplectic structure
of the classical solution space.

Equations \eqref{eq:psiphi_1}-\eqref{eq:psipsi}
allow the mode coefficients to be extracted from
the conserved pairing, giving the reconstruction formula
\begin{align}
 f(t,\sigma)=&
\frac{1}{\pi i}(\psi_0,f)\,\varphi_0(t,\sigma)-\frac{1}{\pi i}(
\varphi_0,f)\,\psi_0(t,\sigma)
\nn
&
+\sum_{n\neq 0}\frac{1}{\pi n}
(\psi_n,f)\,\psi_n(t,\sigma).
\label{eq:fexp}
\end{align}
In this sense the modes form a complete system in the solution space.
Two distributional identities following from this mode system will be
useful below:
\begin{align}
 G_1(\sigma,\sigma')\equiv& \sum_{n\in \mathbb{Z}} \psi_n(t,\sigma)
\psi^*_n(t,\sigma')=\pi\delta(\sigma-\sigma'),
\label{eq:G1}
\\
 G_2(\sigma,\sigma')\equiv &\,i\psi_0(t,\sigma)\varphi_0^*(t,\sigma')
-i\varphi_0(t,\sigma)\psi_0^*(t,\sigma')
\nn
&+\sum_{n\neq 0}\frac{1}{n}\psi_n(t,\sigma)
\psi_n^*(t,\sigma')=0.
\label{eq:G2}
\end{align}
It should be emphasized that these identities are understood in the
sense of distributions rather than as pointwise identities.  Namely,
they acquire their meaning only after being integrated against
sufficiently regular test functions.
In particular,
pointwise endpoint values do not define additional
contributions
to the corresponding distributional kernels.

The canonical momenta $P(t,\sigma)$ and
$\tilde{P}(t,\sigma)$
conjugate to the string coordinates $X(t,\sigma)$ and $\tilde{X}(t,\sigma)$,
respectively, are given by
\begin{align}
P(t,\sigma)=&\frac{1}{2\pi\alpha'}\dot{\tilde{X}}(t,\sigma)
+\frac{i\tan\pi\lambda}{4\pi\alpha'}\{
\delta(\sigma)-\delta(\pi-\sigma)\}\tilde{X}(t,\sigma),
\label{eq:Pdef}
\\
\tilde{P}(t,\sigma)=&
\frac{1}{2\pi\alpha'}\dot{X}(t,\sigma)
-\frac{i\tan \pi\lambda}{4\pi\alpha'}
\{\delta(\sigma)-\delta(\pi-\sigma)\}X(t,\sigma).
\label{eq:tildePdef}
\end{align}
We then impose the canonical commutation relations
\begin{align}
&
 [X(t,\sigma),\,P(t,\sigma')]=i\delta(\sigma-\sigma'),
\\
&
 [\tilde{X}(t,\sigma),\,\tilde{P}(t,\sigma')]=i\delta(\sigma-\sigma'),
%\\
%&[X(t,\sigma),\,\tilde{X}(t,\sigma')]=0,
\end{align}
with all other commutators vanishing.

We expand string coordinates in the basis introduced above:
\begin{align}
 X(t,\sigma)=&\, y_0 \psi_0(t,\sigma)+\sqrt{2\alpha'}\alpha_0\,
\varphi_0(t,\sigma)
\nn
&+i\sqrt{2\alpha'}\sum_{n\neq 0}\frac{1}{n}\alpha_n\,
\psi_n(t,\sigma),
\\
 \tilde{X}(t,\sigma)=&\, \tilde{y}_0 \tilde{\psi}_0(t,\sigma)
+\sqrt{2\alpha'}\tilde{\alpha}_0\,
\tilde{\varphi}_0(t,\sigma)
\nn
&
+i\sqrt{2\alpha'}\sum_{n\neq 0}\frac{1}{n}
\tilde{\alpha}_n\,
\tilde{\psi}_{n}(t,\sigma).
\end{align}
From the form of a generic expansion \eqref{eq:fexp},
the zero modes $y_0$ and $\tilde{y}_0$ and
mode coefficients $\alpha_n$ and
$\tilde{\alpha}_n$ are given by
\begin{align}
& y_0=-\frac{1}{\pi i}(\varphi_0, X),\quad
\alpha_n=\frac{1}{\sqrt{2\alpha'}\pi i}(\psi_n,\,X),
\nn
&
\tilde{y}_0=-\frac{1}{\pi i}(\tilde{\varphi}_0,\tilde{X}),
\quad
\tilde{\alpha}_n=\frac{1}{\sqrt{2\alpha'}\pi i}(\tilde{\psi}_{n},\,\tilde{X}).
\end{align}
In evaluating these pairings, the time
derivatives $\dot{X}$ and $\dot{\tilde{X}}$ appearing
in $DX$ and $\tilde{D}\tilde{X}$
can be expressed in terms of the canonical momenta
$P$ and $\tilde{P}$ using \eqref{eq:Pdef}
and \eqref{eq:tildePdef}~\cite{Abouelsaood:1986gd,Ishibashi:2016xak}.
For example, $y_0$ and $\alpha_n$ can be written as
\begin{align}
 y_0=&-\frac{1}{\pi i}
\int_0^\pi d\sigma\Big[\varphi_0^*(t,\sigma)
\Big\{2\pi i\alpha' \tilde{P}(t,\sigma)
\nn
&
+\frac{\tan\pi\lambda}{2}
\{\delta(\sigma)-\delta(\pi-\sigma)\}X(t,\sigma)
\Big\}-i\psi_0^*(t,\sigma)X(t,\sigma)\Big],
\nn
 \alpha_n=&\frac{1}{\sqrt{2\alpha'}\pi i}
\int_0^\pi d\sigma\psi_n^*(t,\sigma)
\Big[2\pi i\alpha' \tilde{P}(t,\sigma)
\nn
&+\frac{\tan\pi\lambda}{2}
\{\delta(\sigma)-\delta(\pi-\sigma)\}X(t,\sigma)+nX(t,\sigma)
\Big].
\label{eq:alpha_extract}
\end{align}
The canonical commutation relations then
give the commutation relations among modes:
\begin{align}
& [y_0,\,\tilde{\alpha}_0]=i\sqrt{2\alpha'},\quad
[\tilde{y}_0,\,\alpha_0]=i\sqrt{2\alpha'},
\label{eq:yalpha0}\\
&
[\alpha_n,\,\tilde{\alpha}_m]=n\delta_{n+m,0},
\end{align}
with all other commutators vanishing.

Conversely, using these oscillator commutation relations,
we can
reconstruct the canonical commutation relations for
$X$, $\tilde{X}$, $P$ and $\tilde{P}$.
In this calculation, the functions $G_1$ and $G_2$ introduced in
\eqref{eq:G1} and \eqref{eq:G2} naturally appear on the right-hand
sides.
In particular, the commutator between the string coordinates
is proportional to $G_2$,
\begin{align}
[X(t,\sigma),\tilde X(t,\sigma')]=&\,
2\alpha'G_2(\sigma,\sigma')
\nn
=&
\left\{
\begin{array}{ll}
 \pi \alpha' \sin 2\pi \lambda& (\sigma=\sigma'=0)\\
- \pi \alpha' \sin 2\pi \lambda& (\sigma=\sigma'=\pi)\\
 0 & (\mathrm{otherwise}).
\end{array}
\right.
\end{align}
The nonvanishing boundary values reproduce the familiar noncommutativity
of open-string endpoints in
a constant background field~\cite{Seiberg:1999vs}.
Nevertheless, as emphasized above, $G_2$ vanishes as a
distribution, so that the canonical commutation relation
$[X(t,\sigma),\,\tilde{X}(t,\sigma')]=0$ remains valid in the
distributional sense.
Similar endpoint contributions appear in the
commutators involving the canonical momenta, such as $[X,\,P]$ and
$[P,\,\tilde{P}]$.
Although some of these terms contain delta functions in one
of the world-sheet coordinates,
their coefficients, regarded as ordinary functions of the other coordinate,
are supported only at isolated endpoint values.
As distributions defined by locally integrable functions,
these coefficients therefore vanish,
and hence the corresponding terms vanish as distributions 
in the two variables $(\sigma,\sigma')$.
Thus the canonical
commutation relations are to be understood as distributional relations
rather than pointwise identities.
In this sense, despite the
background-dependent boundary conditions and endpoint noncommutativity,
the canonical variables $X,\,\tilde{X},\,P,\,\tilde{P}$ can be
regarded as background-independent universal objects.
This observation will
allow us to relate the mode representations for different
backgrounds through a Bogoliubov transformation.

\section{Bogoliubov transformations between different magnetic
backgrounds}

As shown in the previous section, the canonical variables $X,\,\tilde{X},\,P$
and $\tilde{P}$ are universal, while their mode decomposition
depends on $\lambda$. We therefore compare the modes at two backgrounds
by evaluating the extraction formulas \eqref{eq:alpha_extract}
for $\lambda'$ with the
universal canonical variables expanded in modes at $\lambda$. This
directly gives the Bogoliubov transformation between the two mode
representations.

For our purpose, it is sufficient to consider an infinitesimal change of
the background, $\lambda\rightarrow \lambda'=\lambda+\delta\lambda$.
The essential point is that the background is varied while the universal
canonical variables themselves are kept fixed: 
\begin{align}
 \delta X(t,\sigma)=\delta\tilde{X}(t,\sigma)
=\delta P(t,\sigma)=\delta\tilde{P}(t,\sigma)=0.
\end{align}
The variation therefore acts only on the background-dependent mode
decomposition of these universal variables. 
Accordingly, in varying
\eqref{eq:alpha_extract}, only the explicit $\lambda$ dependence of the
basis functions and of the coefficients appearing in the mode-extraction
formulas has to be taken into account.

To make the structure of the transformation transparent,
we
introduce the Heisenberg-picture mode variables
\begin{align}
 y_0(t)&\equiv
 y_0+\sqrt{2\alpha'}\,\alpha_0\,t,
\\
 \alpha_n(t)&\equiv
 \alpha_n e^{-int}.
\end{align}
Substituting the mode expansions of the universal canonical variables
into the variation of the extraction formulas~\eqref{eq:alpha_extract}
and performing the $\sigma$ integration, we obtain the infinitesimal
transformation of the modes.  Although intermediate expressions contain
explicit powers of $t$, all such terms combine into the
Heisenberg-picture variables $y_0(t)$ and $\alpha_n(t)$.  The resulting
transformation therefore contains no explicit time dependence beyond
that carried by the modes themselves.

The resulting transformations are
\begin{align}
 \delta y_0(t)
=&
\,\pi \delta \lambda \frac{1}{2}\tan \pi \lambda \, y_0(t)
\nn
&
+\delta \lambda \,i\sqrt{2\alpha'}
\sum_{m\neq 0}\frac{1}{4m^2}
\Big[
4\{1-(-1)^m\}
\nn
&+\{1+(-1)^m\}m\pi\tan \pi \lambda\Big]\,\alpha_m(t),
\end{align}
and, for the zero-mode coefficient $\alpha_0(t)$,
\begin{align}
 \delta \alpha_0(t)=&
-\pi \delta \lambda\, \frac{1}{2}\tan\pi \lambda
\,\alpha_0(t)
\nn
&
+\delta \lambda
\sum_{m\neq 0}\frac{1}{2m}\{1-(-1)^m\}\,\alpha_m(t),
\end{align}
while for the nonzero modes $n\neq 0$,
\begin{align}
 \delta \alpha_n(t) =&\delta \lambda \frac{1}{\sqrt{2\alpha'}}
\frac{i}{2}\{1-(-1)^n\}\,y_0(t)
\nn
&
-\delta \lambda\frac{1}{4n}
\Big[4 \{1-(-1)^n\}+\{1+(-1)^n\}
n\pi \tan \pi \lambda\Big]\alpha_0(t)
\nn
&
+
\delta \lambda \sum_{\substack{m\neq 0\\ m\neq n}}
\frac{1}{2m}\frac{m+n}{m-n}\{1-(-1)^{n+m}\}
\alpha_m(t).
\end{align}
The transformations of $\tilde y_0(t)$ and
$\tilde\alpha_n(t)$ are obtained in the same way from the
corresponding mode-extraction formulas and will not be displayed
explicitly.

The transformations in both sectors are generated by a single
operator.  Since their coefficients are independent of $t$, we
may set $t=0$ and define
\begin{align}
 {\cal B}(\lambda)=&
-\pi \frac{\tan\pi \lambda}{2i\sqrt{2\alpha'}}(
\alpha_0\tilde{y}_0+y_0\tilde{\alpha}_0)
\nn
&
-
\sum_{m\neq 0}\frac{1}{4m^2}
\Big[
4\{1-(-1)^m\}
\nn
&
\hspace{1cm}
+\{1+(-1)^m\}m\pi \tan \pi \lambda\Big]
(\alpha_m\tilde{\alpha}_0-\tilde{\alpha}_{-m}\alpha_0)
\nn
&
+\frac{1}{i\sqrt{2\alpha'}}
\sum_{m\neq 0}\frac{1}{2m}\{1-(-1)^m\}(\alpha_m \tilde{y}_0
+\tilde{\alpha}_{-m}y_0)
\nn
&
-
\sum_{\substack{m,n\neq 0\\ m\neq n}}
\frac{1}{2mn}\frac{m+n}{m-n}\{1-(-1)^{n+m}\}
\,\alpha_m \tilde{\alpha}_{-n}.
\label{eq:Bdef}
\end{align}
The generator ${\cal B}(\lambda)$
is determined by the canonical transformation up to a central term.
We adopt the formal anti-Hermiticity prescription
for this ambiguity, leaving only a $\lambda$-dependent
imaginary central term,
which affects only the phase convention of the transformed vacuum.
Here,
anti-Hermiticity is understood formally at the level of the canonical
oscillator algebra.
As shown below, however, this does not imply that
the transformation is unitarily implementable on a common Fock space.
${\cal B}(\lambda)$ generates the infinitesimal Bogoliubov
transformation:
\begin{align}
&
 \delta y_0 =\delta \lambda\,[{\cal B}(\lambda),\,y_0],
\quad\delta \tilde{y}_0 = 
\delta \lambda\,[{\cal B}(\lambda),\,\tilde{y}_0],
\nn
&
 \delta \alpha_n = \delta \lambda\,[{\cal B}(\lambda),\,\alpha_n],
\quad\delta \tilde{\alpha}_n = 
\delta \lambda\,[{\cal B}(\lambda),\,\tilde{\alpha}_n].
\label{eq:Btrans}
\end{align}
These relations give the infinitesimal change of the mode representation
induced by a change of the boundary condition, with the universal
canonical variables held fixed.

\section{Dirac-Born-Infeld  action from the Bogoliubov transformation}

We first consider the Hilbert spaces associated with different
values of the background magnetic field.  Since
$\tilde{X}$ is the Hermitian conjugate
of $X$, the mode operators satisfy
\begin{align}
y_0^\dagger &=\tilde{y}_0,
\qquad
\alpha_n^\dagger = \tilde{\alpha}_{-n}.
\label{eq:hermiticity}
\end{align}
For a fixed value of $\lambda$,
we take $|0\rangle_\lambda$ to be the zero-momentum vacuum,\footnote{
More precisely, one may regard the state space as
a rigged Hilbert space~\cite{delaMadrid:2005qdg},
$\Phi_\lambda\subset {\cal H}_\lambda\subset \Phi_\lambda^*$
with the zero-momentum vacuum belonging to $\Phi_\lambda^*$.
For the treatment of generalized eigenstates and their distributional pairings
in the Schr\"odinger representation, see also \cite{Arisue:1980hk}.
}
corresponding to the $SL(2,\mathbb{R})$-invariant vacuum of the matter sector,
satisfying
\begin{align}
\alpha_n(\lambda)|0\rangle_\lambda
=
\tilde\alpha_n(\lambda)|0\rangle_\lambda=0,
\qquad
(n\geq 0).
\label{eq:vacuum}
\end{align}
The corresponding Fock representation is formally
generated from states of the form
\begin{align}
\alpha_{-n_1}(\lambda)\cdots\alpha_{-n_r}(\lambda)
\tilde\alpha_{-m_1}(\lambda)\cdots\tilde\alpha_{-m_s}(\lambda)
|0\rangle_\lambda 
\quad (n_i,\,m_i>0).
\end{align}

We now consider an infinitesimal change of the background,
$\lambda\to\lambda+\delta\lambda$. 
Following the prescription of Kugo and
Zwiebach~\cite{Kugo:1992md}, we formally relate the vacua associated
with different backgrounds by transforming them
through the infinitesimal Bogoliubov transformation:
\begin{align}
|0\rangle_{\lambda+\delta\lambda}
=
|0\rangle_\lambda
+\delta\lambda\,{\cal B}(\lambda)\,|0\rangle_\lambda
+O(\delta\lambda^2).
\label{eq:formalvac}
\end{align}
This relation should not be interpreted as a unitary transformation
within a single Fock space. Indeed, as we show below, the infinitesimal
variation cannot be represented as a vector
in the original Fock representation.

The problem becomes apparent 
already from the oscillator contribution to the formal norm of the first-order variation.
Acting with ${\cal B}(\lambda)$ on the vacuum, we obtain
\begin{align}
&
{\cal B}(\lambda)|0\rangle_\lambda
=
\Bigg[\frac{\pi}{2}\tan \pi \lambda
\nn
&
-\frac{1}{\sqrt{2\alpha'}i}
\sum_{m=1}^\infty
\frac{1}{2m}\{1-(-1)^m\}(\alpha_{-m}\tilde{y}_0
-\tilde{\alpha}_{-m}y_0)
\nn
&
+
\sum_{m\geq 1,n\geq 1}
\frac{1}{2mn}\frac{m-n}{m+n}\{1-(-1)^{n+m}\}
\,\alpha_{-m} \tilde{\alpha}_{-n}\Bigg]\ket{0}_\lambda.
\label{eq:B|0>}
\end{align}
Using the formal conjugation rule of the oscillator algebra,
the corresponding bra expression is
\begin{align}
-&{}_\lambda\langle0|{\cal B}(\lambda)=
{}_\lambda\langle0|\,
\Bigg[\frac{\pi}{2}\tan \pi \lambda
\nn
&
+\frac{1}{\sqrt{2\alpha'}i}
\sum_{m=1}^\infty
\frac{1}{2m}\{1-(-1)^m\}(\tilde{\alpha}_{m}y_0
-\alpha_{m}\tilde{y}_0)
\nn
&
+
\sum_{m\geq 1,n\geq 1}
\frac{1}{2mn}\frac{m-n}{m+n}\{1-(-1)^{n+m}\}
\,\tilde{\alpha}_{m} \alpha_{n}\Bigg].
\label{eq:-<0|B}
\end{align}
The formal norm of the first-order vacuum variation then diverges:
\begin{align}
&\lVert\,{\cal B}(\lambda)|0\rangle_\lambda\,\rVert^2
\geq
\sum_{m\geq 1,\,n\geq 1}
\frac{1}{2mn}\left(
\frac{m-n}{m+n}\right)^2\{1-(-1)^{n+m}\}
\,{}_\lambda \langle 0|0\rangle_\lambda.%=\infty.
\label{eq:divnorm}
\end{align}
The purely oscillator contribution
of \eqref{eq:divnorm}
already diverges by itself.
Equivalently, the off-diagonal part of the Bogoliubov
transformation, which mixes creation and annihilation operators,
does not satisfy the Hilbert-Schmidt condition. 

The transformation therefore cannot be implemented by a unitary operator
on ${\cal H}_\lambda$,
even though its action on each individual mode is well
defined. Consequently, the Fock representations corresponding to
different
values of $\lambda$ are unitarily inequivalent.

The unitary inequivalence does not, however, make the Bogoliubov
transformation physically irrelevant.
Although the vacua cannot be identified
as vectors in a common Fock space, 
their relative normalization may still carry
physical information.

The annihilation conditions~\eqref{eq:vacuum} do not fix the overall normalization
of the zero-momentum vacuum. We therefore investigate whether the
background dependence of this normalization,
\begin{align}
{\cal N}(\lambda)\equiv
{}_\lambda\langle0|0\rangle_\lambda,
\label{eq:vacnorm}
\end{align}
as determined by the
Bogoliubov transformation, reproduces the relative boundary
$g$-function
known independently from BCFT.

From \eqref{eq:formalvac}, the
formal variation of the ket vacuum is
\begin{align}
 \frac{d}{d\lambda}|0\rangle_\lambda&=
{\cal B}(\lambda)|0\rangle_\lambda.
\end{align}
Since the infinitesimal transformation is not
unitarily implementable, this equation is understood
as a formal transform relation between different
representations, rather than as an operator identity
on a common Hilbert space.
The corresponding variation of the bra vacuum
is obtained independently from the conjugate
Bogoliubov transformation:
\begin{align}
 \frac{d}{d\lambda}{}_\lambda\langle0|
&=
-{}_\lambda\langle0|{\cal B}(\lambda),
\label{eq:d<0|}
\end{align}
as given in \eqref{eq:-<0|B}.
It then follows that\footnote{
\eqref{eq:d<0|} represents the formal
algebraic conjugation of the Bogoliubov transformation.
Actually, the variation of the vacuum normalization
should be evaluated using an infrared-regularized pairing,
as explained in~\ref{app:zero-mode}.}
\begin{align}
\frac{d}{d\lambda}{\cal N}(\lambda)
=&\left(\frac{d}{d\lambda}{}_\lambda\langle 0|\right)
|0\rangle_\lambda
 +
{}_\lambda\langle 0|
\left(\frac{d}{d\lambda}|0\rangle_\lambda\right)
\nn
=&\pi \tan \pi \lambda\,{\cal N}(\lambda),
\label{eq:vacnormdiff}
\end{align}
where only the vacuum components in Eqs.~\eqref{eq:B|0>}
and \eqref{eq:-<0|B} contribute, while
the remaining terms vanish by the annihilation conditions~\eqref{eq:vacuum}.

Although ${\cal B}(\lambda)$ is formally anti-Hermitian,
\eqref{eq:divnorm} shows that
it does not define an
infinitesimal unitary transformation within the
original Fock space.
Moreover,
the zero-momentum vacuum itself is a generalized state
rather than a normalizable vector in ${\cal H}_\lambda$.
Therefore, the usual Hilbert-space argument for norm preservation does not
apply to this formal transform.

Integrating~\eqref{eq:vacnormdiff} with
$\lambda=0$ as the reference background, we obtain
\begin{align}
\frac{{\cal N}(\lambda)}{{\cal N}(0)}
=\frac{g(\lambda)}{g(0)}
=\frac{1}{\cos\pi\lambda}
=\sqrt{1+(2\pi\alpha' F_{12})^2}.
\label{eq:DBI}
\end{align}
This reproduces the relative boundary $g$-function,
or equivalently the magnetic-field-dependent part of
the disk partition function.
For a general constant magnetic field, the contributions
from independent two-dimensional blocks combine to
give the DBI factor $\sqrt{\det(1+2\pi\alpha' F)}$.

\section{Concluding remarks}

In this letter, we have shown that the canonical variables of an open
string in a constant magnetic field can be regarded as
background-independent universal objects, with the background dependence
encoded in their mode representations.
The representations for different
magnetic fields are unitarily inequivalent, but the Bogoliubov
transformation nevertheless determines the relative
normalization of their vacua.
The resulting relative vacuum normalization reproduces
the relative boundary $g$-function,
or equivalently the magnetic-field dependence of the disk partition function,
and hence the DBI factor.

The result provides a complementary interpretation of our previous
calculation in~\cite{Ishibashi:2016xak}.
There, the same $g$-function was obtained from the factorization of correlation
functions of boundary-condition-changing operators.
Here, without evaluating such world-sheet correlation functions,
its background dependence is recovered from the Bogoliubov transformation
between inequivalent representations of the universal canonical variables.
The agreement shows that the change of the disk normalization
associated with a change of boundary condition is encoded directly
in the transformation between representations
of the canonical open-string algebra.

This viewpoint suggests several directions for further study.
An immediate question is whether the construction can be
extended to more general background deformations.
In particular, including changes of the space-time metric together
with the antisymmetric tensor background would test
how the vacuum normalization and
the open-string coupling combine to reconstruct
the corresponding disk normalization.

If closed-string background data can be encoded in this way in the
representation structure of the open-string state space,
such a framework may provide a string-theoretic realization of the
holographic idea that closed-string geometry emerges from
open-string degrees of freedom.

It is also interesting to reconsider from this viewpoint the classical
open string field theory solution describing a constant magnetic
background~\cite{Ishibashi:2016xak}.  Although the Fock representations
for different magnetic backgrounds are unitarily inequivalent, the
corresponding classical solution is a well-defined state in the original
open-string Fock space~\cite{Ishibashi:2018ynb}.  Understanding how this
string field theoretic description is related to the Bogoliubov
transformation studied here may clarify how background changes are encoded in
string field theory.

\appendix
\section{Infrared regularization of the zero-mode pairing}
\label{app:zero-mode}

We clarify the zero-mode contribution to the vacuum-normalization
equation~\eqref{eq:vacnormdiff}.
The zero-momentum vacuum in~\eqref{eq:vacuum} is a
generalized eigenstate, and 
its matrix elements cannot be treated as
ordinary Hilbert-space matrix elements without a prescription. The
distributional nature of generalized eigenstates and their pairings has
been emphasized in the generalized Schrödinger representation of Arisue
et al.  \cite{Arisue:1980hk}.
We therefore introduce an infrared regularization to
examine the apparent inconsistency between the left and right actions of
the zero-mode generator.

Consider a real canonical pair
$[x,\,p]=i$, with $p=-i\partial_x$, and the symmetric dilatation operator
\begin{align}
{\cal D}=xp+px=-2ix\partial_x-i.
\end{align}
The zero-momentum state $|0\rangle$, represented by a constant
wavefunction, satisfies $p|0\rangle=0$.
Nevertheless, the canonical
commutation relation implies
$xp|0\rangle=0,\ px|0\rangle=-i|0\rangle$,
and hence
\begin{align}
{\cal D}|0\rangle=-i|0\rangle. 
\end{align}
This is consistent with the Hermiticity of ${\cal D}$
on its Hilbert-space domain,
since $|0\rangle$ is not a normalizable vector in $L^2(\mathbb R,dx)$.

Introduce a smooth positive weight $\mu_L(x)$ with an infrared cutoff $L$,
satisfying $\int_{\mathbb R}dx\,\mu_L(x)<\infty$
and $x\mu_L(x)\to0$ as $x\to\pm\infty$, and define
\begin{align}
\langle f|g\rangle_L
=\int_{\mathbb R}dx\,\mu_L(x)f(x)^*g(x),
\qquad
V_L=\langle0|0\rangle_L.
\end{align}
Integration by parts gives the adjoint of ${\cal D}$ with respect to this pairing:
\begin{align}
{\cal D}_L^\dagger
={\cal D}-2ix\frac{\mu_L'(x)}{\mu_L(x)}.
\end{align}
Consequently,
\begin{align}
\frac{\langle0|i{\cal D}|0\rangle_L
-\langle0|i{\cal D}_L^\dagger|0\rangle_L}{V_L}
&=-\frac{2}{V_L}\int_{\mathbb R}dx\,x\mu_L'(x)
=2. 
\label{eq:DD2}
\end{align}

This identity holds at finite $L$.
For example, with the Gaussian regulator $\mu_L(x)=e^{-x^2/L^2}$,
one obtains $V_L=\sqrt{\pi}L$ and ${\cal D}_L^\dagger={\cal D}+4ix^2/L^2$.
Although ${\cal D}_L^\dagger-{\cal D}$ vanishes pointwise at fixed $x$ as $L\to \infty$,
its normalized matrix element in \eqref{eq:DD2} remains finite.

The zero-mode part of the Bogoliubov generator \eqref{eq:Bdef} is
\begin{align}
{\cal B}_0(\lambda)
=-\frac{\pi\tan(\pi\lambda)}{2i\sqrt{2\alpha'}}
(\alpha_0\tilde y_0+y_0\tilde\alpha_0).
\end{align}
Using the commutation relations \eqref{eq:yalpha0},
it can be expressed in terms of two real canonical pairs as
\begin{align}
{\cal B}_0(\lambda)
=\frac{i\pi\tan(\pi\lambda)}4({\cal D}_1+{\cal D}_2),
\qquad {\cal D}_a=x_ap_a+p_ax_a,
\end{align}
where $x_1=(y_0+\tilde{y}_0)/\sqrt{2}$,
$x_2=(y_0-\tilde{y}_0)/(\sqrt{2}i)$,
$p_1=(\alpha_0+\tilde{\alpha}_0)/(2\sqrt{\alpha'})$,
$p_2=(\alpha_0-\tilde{\alpha}_0)/(2i\sqrt{\alpha'})$.
Applying \eqref{eq:DD2} to both real directions gives
\begin{align}
\frac{\langle0|({\cal B}_0+{\cal B}_{0,L}^\dagger)|0\rangle_L}
{\langle0|0\rangle_L}
=\pi\tan(\pi\lambda). 
\label{eq:<BB0L>}
\end{align}
Here ${\cal B}_{0,L}^\dagger$ denotes
the adjoint with respect to the regularized pairing,
rather than the formal algebraic replacement ${\cal B}_0\to-{\cal B}_0$.
The result is independent of the regulator profile
within the stated class and survives the removal of the infrared regulator.
It provides a regularized interpretation of
the finite zero-mode contribution appearing in~\eqref{eq:vacnormdiff}.

%%%%%%%%%%%%%%%%%%%%%%%%%%%%%%%%%%%%%%%
\section*{Acknowledgments}
The initial motivation for this work arose from an earlier collaboration
with Y.~Satoh. We would like to thank T.~Kugo and Y.~Satoh for valuable
comments that helped clarify the physical interpretation of our
results, and H.~Itoyama, S.~Seki and M.~Yamada for useful discussions
and comments.  The authors thank the Yukawa Institute for Theoretical
Physics at Kyoto University for its hospitality.
Discussions during the YITP workshop
YITP-W-26-08 on
``Strings and Fields 2026''
were helpful in completing this
work.  T.~T.~was supported in part by JSPS KAKENHI Grant Numbers
\#23K03388 and \#26K22332.  A.~T.~was supported by the Initiative for
Realizing Diversity in the Research Environment.
%%%%%%%%%%%%%%%%%%%%%%%%%%%%%%%%%%%%%%%%%%%%%%%%%%%%%%%
%
%\section*{Declaration of generative AI and AI-assisted technologies
%in the manuscript preparation process}
%
%During the preparation of this work the authors used ChatGPT in order
%to improve the language and readability of the manuscript.
%After using this tool/service, the authors reviewed and edited the
%content as needed and take full responsibility for the content of
%the published article.
%
%
%%%%%%%%%%%%%%%%%%%%%%%%%%%%%%%%%%%%%%%%%%%%%%%%%%%%%

\end{document}